\documentclass[prb,twocolumn,groupedaddress]{revtex4}

\usepackage{amssymb}

\usepackage{bm}

\usepackage{amsmath}

\newcommand{\ic}{{I_{\rm c}}}

\input{epsf}

\begin{document}

\title{Theory of anomalous magnetic interference pattern in mesoscopic
superconducting/normal/superconducting Josephson junctions}
\author{Daniel E.~Sheehy\rlap,$^1$ Alexandre M. Zagoskin$^{1,2}$}
\affiliation{ $^1$Department of Physics and Astronomy, University
of British Columbia,
6224 Agricultural Road, Vancouver, B.C.~V6T1Z1, Canada\\
$^2$D-Wave Systems Inc., 320-1985 W.~Broadway, Vancouver, B.C.,
V6J 4Y3, Canada}
\date{January 27, 2003}
\begin{abstract}



The magnetic interference pattern in mesoscopic SNS Josephson
junctions is sensitive to the scattering in the normal part of the
system. In this paper we investigate it, generalizing Ishii's
formula for current-phase dependence to the case of normal
scattering at NS boundaries in an SNS junction of finite width.
The resulting flattening of the first diffraction peak is
consistent with experimental data for
S-2DEG-S mesoscopic junctions.


\end{abstract}

\maketitle





\section{Introduction}

The Fraunhofer diffraction pattern characterizing the dependence
of the critical Josephson current on an applied external magnetic
flux is a well-known feature of tunneling or dirty
superconducting-normal-superconducting (SNS)
junctions~\cite{REF:Tinkham}.  The latter is defined by the
condition that quasiparticle scattering length by impurities in
the normal layer, $l_i$, is much less than the characteristic
dimensions  ($W,L$) of this layer. For the case of clean SNS
junctions, this pattern is more complex and demonstrates a
distinct triangular central peak~\cite{REF:Svidzinskii}. The
pattern is $\Phi_0$-periodic in either case.
More recent experimental and theoretical efforts have revealed,
that in mesoscopic SNS junctions this pattern may be strongly
altered by geometrical effects. For example, Heida et
al~\cite{REF:Heida} observed a $2\Phi_0$-periodic diffraction
pattern in narrow junctions, in contrast to the expected $\Phi_0$.
The essential features of this behavior were explained
theoretically~\cite{REF:Barzykin,REF:Ledermann} using a
semiclassical technique that accounts for the restricted geometry
in terms of the classical trajectories associated with
current-carrying Andreev states states in the normal layer.
In this paper we generalize this approach to include the effects
of normal reflection at the NS boundaries, which are treated in the
Blonder-Tinkham-Klapwijk (BTK) approximation~\cite{REF:BTK}. The
motivation for this work was provided by the JHAT (Jensen, Harada,
Akazaki, and Takayanagi) experiment~\cite{REF:Takayanagi}. They
observed critical current oscillations vs. flux in a wide
ballistic S-2DEG-S junction, which exhibited apparently random
small-scale (a fraction of $\Phi_{0}$) oscillations on top of a
flattened first interference peak. As the first step towards
understanding this data, here we extend previous
work~\cite{REF:Barzykin,REF:Ledermann} (which has focused mainly
on the case without normal reflection at the NS boundary) by including
closed trajectories formed by reflection from the walls of the
device as well as from the NS boundaries. Such trajectories may be
important because, for them,  the effective area penetrated by the
magnetic field (and therefore the field-dependent phase of the
corresponding contribution to the supercurrent) change smoothly
with their variations.  By contrast, the contribution due to
chaotic trajectories is expected to exhibit a rapid change with
variation of the trajectory  and thus the field dependence
effectively cancels on scales below 
 $\Phi _{0}$, like in an SNS junction
in the diffusive limit. We show, that such an  approach captures some,
but not all, essential features of the experiment.

\begin{figure}[hbt]
\par
\epsfxsize=2.9in
\par
\par
\centerline{\epsfbox{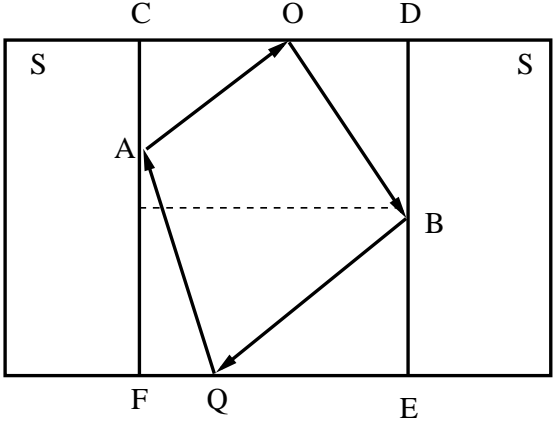}}
\par
\vskip0.50cm \caption{Schematic picture of a closed quasiparticle
trajectory carrying supercurrent between the superconducting leads
(labelled S). The dashed line is at $y = 0$. } \label{fig:one}
\end{figure}

The general idea behind the quasiclassical approach is to
represent the Josephson current through the SNS junction as a sum
of contributions from various quasiclassical trajectories linking
the superconductors~\cite{REF:Barzykin}. Such a generalization of
Ishii's formula~\cite{REF:Ishii} follows from the quasiclassical
Eilenberger equations and gives a transparent description suitable
for complicated geometries. We will assume that the conditions of
its applicability ($\lambda_F\ll\xi_0\ll W,L$, where $\lambda_F$
is the Fermi wavelength and $\xi_0$ is the superconducting
coherence length [\lq\lq Cooper pair size\rq\rq]) are satisfied.
Our approach is to first calculate the density  $N(\xi)$ of
Andreev states in the normal area and then express the Josephson
current through it. (Here $\xi$ is the quasiparticle energy
measured from the Fermi level.)
For simplicity, we consider the case of a rectangular normal
region with dimensions $W\times L$ (see Fig.~\ref{fig:one}), with
specular normal reflections from the boundaries (given by the line
segments $CD$ and $FE$ in Fig.~\ref{fig:one}).
%
%

%
The flux $\Phi=HWL$ produced by the magnetic field $\vec{
H}=H\vec{e}_z $ through the normal region can be described by the
vector potential
\begin{equation}
\label{eq:one}
\vec{A}=-Hy\cdot \vec{e}_{x}.
\end{equation}
The applied magnetic field is screened by the superconducting regions,
producing screening currents over a length given by $\lambda$, the 
magnetic penetration depth.  As is standard~\cite{REF:Tinkham}, we 
shall neglect the variation of the superconducting phase and magnetic field
over this length scale; thus, our results apply at $W,L \gg \lambda$.


Before turning to the calculation of the current, let us briefly
remark on why closed trajectories are of principal importance.
The introduction of normal reflection on NS boundary takes away
the simplicity of quasiclassical trajectories with Andreev
reflections only (when the hole retraces the electron's path).
Even if assume (as we do), that the Andreev reflection exactly
reverts the group velocity of a quasiparticle, and that a normal
reflection is specular, the quasiparticle trajectory will be
generally open, with multiple reflection points at an NS boundary.
Each SNS leg of such a trajectory coherently contributes to the
latter's partial Josephson current $I_{tr.}$, with wildly varying
phase factors. Therefore the total contribution of such
trajectories will be suppressed. The exceptions are closed
trajectories, where phase gains are systematic. Such trajectories
contribute most to the quasiclassical density of
states~\cite{REF:Adagideli}.
If normal reflection amplitude from NS boundary is small,
$|\mathcal{R}|\ll 1$, the leading contribution will be from the
trajectories with the minimum possible number of  reflections on 
the NS
boundaries (one each, see Fig.~\ref{fig:one}) (class A trajectories),
which simplifies the situation. On the other hand, in this case we
can not discard the contributions from open (class B)
trajectories, since only the first leg will significantly
contribute to the current, and the above \lq\lq phase cancelling"
argument no longer applies.

In the following, we assume that 1) Andreev reflection on NS 
boundary prevails and 2) the scattering in the normal region 
is negligible~\cite{note:scattering}.
We will therefore concentrate on the two above classes of
trajectories. Consider then the closed trajectory AOBQ in
Fig.~\ref{fig:one}, where electrons propagate clockwise, and
holes, related to the electrons by Andreev reflections from NS
boundaries (at A and B in Fig.~\ref{fig:one}), propagate
counterclockwise. (The results presented here are valid for a
trajectory with an arbitrary number of reflections from the side
walls). The phase gain for an electron from the magnetic field
along AOB is simply related to the flux enclosed by the trajectory
and the dashed line $y=0$. We introduce the phase gains 
$w$ and $\bar{w}$ along AOB and BQA, via
\begin{eqnarray}
w\cdot \Phi _{0}&=&\int_{AOB}\overrightarrow{A}\cdot
d\overrightarrow{l} =-H\int_{0}^{L}y(x)dx,
\nonumber \\
&=&-H(S_{AOBEF}-WL/2),  \label{eq:phase1} \\
\overline{w}\cdot \Phi _{0}&=&\int_{BQA}\overrightarrow{A}\cdot d
\overrightarrow{l}=-H\int_{L}^{0}y(x)dx,
\nonumber \\
&=&+H(S_{BQAFE}-WL/2). \label{eq:phase2}
\end{eqnarray}

We have expressed the appropriate area in Eq.~(\ref{eq:phase1})
[and similarly in Eq.~(\ref {eq:phase2})] as the difference of the
area below AOB (which we denote by $S_{AOBEF}$) and the area
below the dashed line. Our expression for the Josephson current
carried by a particular trajectory will involve both the sum, $\nu
\equiv w+\overline{w}$, and difference, $\Omega \equiv
w-\overline{w}$, of these phases:
\begin{eqnarray}
\nu \cdot \Phi _{0}&=& -H(S_{AOBEF}-S_{BQAFE}), \label{eq_nu}
\\
&=&-HS_{AOBQ}\equiv -\Phi_{ex},
\label{eq_nu1} \\
\Omega \cdot \Phi _{0}&=&-H(S_{AOBEF}+S_{BQAFE}) + HWL,
\label{eq_Omega}
\end{eqnarray}
where $\Phi_{ex}$ is the flux of the external field through the
loop $S_{AOBQ}$. Let us compute the phase gains (due to the field
and from propagation) accumulated by quasiparticles travelling
along $AOB$ and $AQB$. Denoting the electron (hole) momentum at
energy $E$ by $k$ ($q$), and setting $|AOB|=l$,
$|BQA|=\overline{l}$, we can write for the electron (hole) wave
function the following expressions for the phase gains along
trajectories:
\begin{equation}
\begin{array}{c}
\psi _{e}(A)\overset{AOB}{\leftrightarrows }\psi _{e}(A)e^{i(kl+\pi w)}, \\
\psi _{h}(A)\overset{AOB}{\leftrightarrows }\psi _{h}(A)e^{i(ql-\pi w)}, \\
\psi _{e}(B)e^{i(k\overline{l}+\pi \overline{w})}\overset{AQB}{
\leftrightarrows }\psi _{e}(B), \\
\psi _{h}(B)e^{i(q\overline{l}-\pi \overline{w})}\overset{AQB}{
\leftrightarrows }\psi _{h}(B).
\end{array}
\end{equation}

Due to the proximity effect, the quasiparticle wave functions in
the normal region are composed of electron and hole components,
which are mixed by Andreev reflections on NS boundaries. Thus, we
have
\begin{eqnarray}
\Psi _{AOB}(x)=a\psi _{e}^{\Longrightarrow }(x)+b\psi
_{h}^{\Longleftarrow}(x), \\
\Psi _{BQA}(x)=\overline{a}\psi _{e}^{\Longleftarrow
}(x)+\overline{b}\psi _{h}^{\Longrightarrow }(x).
\end{eqnarray}
Here, the arrows indicate the direction of the group velocity
projection on the x-axis. The amplitudes of normal and Andreev
reflection are denoted by $\mathcal{R}$ and $\mathcal{A}
=-i|\mathcal{A}|e^{\pm i\chi }$, respectively.

For the phase of the latter we have taken the result that is valid
for a \textit{clean} NS boundary and for a quasiparticle\textit{\
exactly }on the Fermi surface ($\xi \equiv E-E_{F}=0).$
The unitarity conditions, which encode the notion that, e.g., an
electron propagating along $BQA$ must undergo either Andreev or
normal reflection at $A$, can be written as:
\begin{eqnarray}
&&\overline{a}\psi_{e}(B)e^{i(k\overline{l}+\pi
\overline{w})}\!=\!\mathcal{R} a\psi_{e}(A)\!-\!i|\mathcal{A}|e^{-i\chi _{A}} \bar{b}\psi_{h}(B)e^{i(q
\overline{l}-\pi \overline{w})}, \nonumber
\\
&&\quad b\psi_{h}(A)=\mathcal{R}\overline{b}\psi_{h}(B)e^{i(q\overline{l}-\pi
\overline{w})}-i|\mathcal{A}|e^{i\chi_{A}} a\psi_{e}(A),
\nonumber
 \\
&&a\psi_{e}(A)e^{i(kl+\pi w)}\!=\!\mathcal{R}\overline{a}\psi_{e}(B) \!-\!i|\mathcal{A}|e^{-i\chi _{B}}
 b\psi_{h}(A)e^{i(ql-\pi w)}, \nonumber
\\
&& \quad \overline{b}\psi _{h}(B)=\mathcal{R}b\psi_{h}(A)e^{i(ql-\pi w)} 
-i|\mathcal{A}|e^{i\chi_{B}}
\overline{a}\psi_{e}(B). \label{eq:unitarity}
\end{eqnarray}
In the next section, we shall use Eq.~(\ref{eq:unitarity}) to find
the density of Josephson current-carrying levels, which 
will yield the Josephson current~\cite{REF:zagoskinbook}.

\section{Dispersion law}
In the above system of equations, we can (up to normalization and immaterial
phase factors) assume $\psi _{e}(A)=\psi _{h}(A)=\psi _{e}(B)=\psi
_{h}(B)\simeq1,$ which leaves us with a homogeneous system of
linear equations for the quantities $\{a,\overline{a},b,
\overline{b}\},$ with matrix
\begin{widetext}
\begin{equation}
M=\left[
\begin{array}{cccc}
\mathcal{R} & -e^{i(k\overline{l}+\pi \overline{w})} & 0 &
-i|\mathcal{A}
|e^{-i\chi _{A}}e^{i(q\overline{l}-\pi \overline{w})} \\
-i|\mathcal{A}|e^{i\chi _{A}} & 0 & -1 &
\mathcal{R}e^{i(q\overline{l}-\pi
\overline{w})} \\
-e^{i(kl+\pi w)} & \mathcal{R} & -i|\mathcal{A}|e^{-i\chi
_{B}}e^{i(ql-\pi
w)} & 0 \\
0 & -i|\mathcal{A}|e^{i\chi _{B}} & \mathcal{R}e^{i(ql-\pi w)} &
-1
\end{array}
\right] .
\label{eq:matrix}
\end{equation}
\end{widetext}

The dispersion law for the supercurrent-carrying quasiparticles
follows from the solvability condition, det
$M=0.$ Henceforth, we explicitly take $l = \bar{l}$, which
is true in case of specular normal reflection and a rectangular
normal region. Introducing $\chi \equiv \chi_{B}-\chi_{A}$, we
have
\begin{eqnarray}
&&\det M=2\mathrm{e}^{i(k+q)l} \cdot D, \\
&&D \equiv \cos\left(\nu\pi + l(k-q)\right) + |\mathcal{A}|^2
\cos\left(\chi - \Omega \pi \right)
\nonumber \\
&& \qquad \qquad -|\mathcal{R}|^2 \cos(k+q)l. \label{eq:ddef}
\end{eqnarray}

At this point we proceed by making an assumption about the
energy-dependent amplitudes for Andreev and normal reflection. As
we are interested in the possibility of normal reflection at the
interface, we take these from  the results of BTK.
Thus, at low energies, we have~\cite{REF:BTK}
\begin{eqnarray}
\label{eq:adef} \mathcal{A} &=& \frac{-i\Delta
\mathrm{e}^{-i\chi}}{2Z\sqrt{\Delta^2-E^2}},
\\
\mathcal{R} &=& \sqrt{1-|\mathcal{A}|^2}.
\end{eqnarray}
The momentum of an electron (hole) with energy $\xi $ is $k (\pm \xi )$;
thus we introduce the dimensionless quantity $\varphi (\xi )=k(\xi )\mathcal{%
L }/2=\sqrt{2m(E_{F}+\xi )}\cdot l$ and define
%
%
$\phi _{\pm }(\xi )\equiv \varphi (\xi )\pm \varphi (-\xi )$,
giving
\begin{eqnarray}
&&D(\xi ,\nu ) = \cos(\phi_{-} +\pi\nu) +
\frac{\Delta^2}{4Z^2(\Delta^2 - \xi^2 )}\cos(\chi - \pi\Omega)
\nonumber
\\
&&\qquad +\frac{\Delta^2}{4Z^2(\Delta^2 - \xi^2)}\cos \phi_+ -
\cos \phi_+.
\end{eqnarray}

In order to calculate the {\bf density of states}, we will use the
approach due to Slutskin~\cite{REF:Slutskin} (applied to SNS
systems in Ref.~\onlinecite{REF:Blom}) to extract the density of
current-carrying states $N(\xi )$ directly
 from the dispersion law.  Thus, $N(\xi )=\sum_{s}\delta (\xi -\xi_{s}),$ where $\xi
_{s}$ are given by the solutions of dispersion equation $D(\xi
,\nu )=0$. A simple variable change yields
\begin{eqnarray}
 \label{eq:appear}
&&N(\xi )=\left| \frac{\partial D(\xi ,\nu )}{\partial \xi
}\right| \delta \left( D(\xi ,\nu )\right)
\\
%
&&\frac{\partial D(\xi ,\nu )}{\partial \xi } \approx
 -\frac{2l}{v_{\mathrm{F}}}\sin(\phi_-+\pi\nu),
 \label{eq:appear2}
\end{eqnarray}
where in Eq.~(\ref{eq:appear2}) we have displayed the low-energy
limit of $\partial D(\xi ,\nu )/\partial \xi$ and approximated
$\partial\phi_{-}/\partial \xi \sim 2l/v_{\mathrm{F}}$ and
$\partial\phi_{+}/\partial \xi \sim 0$.

The delta-function constraint in Eq.~(\ref{eq:appear}) implies
that
\begin{eqnarray}
\label{eq:phasedif} &&\phi_- + \pi\nu + (2m +1 )\pi
 \\
&& \quad = \pm \arccos\left( \frac{\Delta^2}{4Z^2(\Delta^2-\xi^2)}
[\cos(\chi-\pi\Omega)+1]-1\right) \nonumber
\end{eqnarray}
with $m$ being an integer. Let us briefly pause to discuss
(qualitatively) the physical meaning of  Eq.~(\ref{eq:phasedif}).
The argument of the $\arccos$ in Eq.~(\ref{eq:phasedif}) has the
interpretation of a generalized phase difference across the
junction. Since the current is given by differentiating the free
energy with respect to the phase
difference~\cite{REF:zagoskinbook}, we see that the $ \pm $ in
Eq.~(\ref{eq:phasedif}) correspond to levels that carry current in
opposite directions.  We shall refer to these two distinct
contributions to $N(\xi)$ as $N_{\pm}(\xi)$.
The prefactor $|\partial D/\partial \xi| \approx
\frac{2l}{v_{\mathrm{F}}} |\sin(\phi_- +\pi\nu)|$ assigns a
particular weight to the $\delta$-function contributions to
$N_{\pm}$; a direct application of the Poisson summation
formula~\cite{note:poisson} yields (taking the low-energy limit)
\begin{eqnarray}
N(\xi ) &=& N_+(\xi) + N_-(\xi) \label{eq_N}\\
N_\pm(\xi)&\equiv & \sum_{n=-\infty }^{\infty
}(-1)^{n}\frac{l}{\pi v_{ \mathrm{F}}} e^{i\pi \nu n}e^{in\phi
_{-}(\xi )\mp in\Gamma (\chi -\pi\Omega )} \nonumber
\\
\Gamma(\chi)& \equiv &\arccos\left( \frac{1}{4Z^2}[%
\cos\chi+1]-1\right), \label{eq:gamma}
\end{eqnarray}
where we have have rescaled the barrier height $Z$ via
$Z\rightarrow Z\Delta$ and we choose a branch of the $\arccos$ in
Eq.~(\ref{eq:gamma}) such that $\Gamma$ smoothly goes to
$\chi-\pi\Omega$ for $Z\rightarrow 1/2$.
In the next section, we apply Eq.~(\ref{eq:gamma}) to the
calculation of the current due to a single trajectory, which (as
we have briefly noted above) is related to the difference
$N_+(\xi) - N_-(\xi) $.

\section{Josephson current due to a particular trajectory}
Having determined the density of states $N(\xi)$ associated with a
particular trajectory, we now turn to the computation of the
associated current. In the limit of zero normal reflectance the
two terms $N_{\pm}(\xi)$ in Eq.~(\ref{eq:gamma}) correspond to two
separate groups of Andreev levels (AOB and BQA) which carry
current in opposite directions and also move in opposite
directions as  $\chi$  is changed \cite{REF:Bardeen}. The
simplicity of this expression is deceptive: the length $l$ and
phase gains $\nu,\:\Omega$  depend nontrivially on the shape of the
trajectory. The associated contribution to the Josephson current
is given in terms of $N_{\pm}(\xi)$ as~\cite{REF:zagoskinbook}
\begin{equation}
\label{eq:traj}
I_{\mathrm{tr.}} = \frac{-e}{l}\int_{-\infty }^{\infty }d\xi \tanh \left( \frac{%
\beta \xi }{2} \right) \left( N_{+}(\xi )-N_{-}(\xi )\right) v_{Fx} ,\\
\end{equation}
where $e$ is the electric charge; henceforth we shall set $e=1$.
In Eq.~(\ref{eq:traj}), the subscript \lq\lq tr.\rq\rq denotes
that this is the contribution due to one particular trajectory,
such as that pictured in Fig.~\ref{fig:one}. At low energies, $
\phi _{-}(\xi )\approx 2\xi l/v_{\textrm{F}}$.  Utilizing the
integral formula
\begin{equation}
\int_{-\infty }^{\infty }d\xi \tanh \left( \frac{\beta \xi
}{2}\right) e^{in2\xi l /v_{\mathrm{F}}\ }= \frac{2\pi i\left(
1-\delta _{n,0}\right) }{ \beta \sinh \left( 2\pi l n/\left(
\beta v_{\mathrm{F}}\right) \right) }, \nonumber
\end{equation}
we find for the current
\begin{equation}
I_{\mathrm{tr.}} =\sum_{n=1}^{\infty}\frac{ 8\cos \left( \pi \nu
n\right) v_{Fx}}{\beta v_{\mathrm{F}} \sinh \left( 2\pi l
n/\left( \beta v_{\mathrm{F}}\right) \right) } (-1)^{n+1}\sin
n\Gamma (\chi-\pi\Omega ).  \label{eq:currentfinal}
\end{equation}
Equation~(\ref{eq:currentfinal}), together with Eqs.(\ref{eq_N}),
(\ref{eq:gamma}), is the central result of this paper~\cite{note:gauge}.

For closed-loop trajectories we must take into account that there
always will be a related contribution from the same trajectory,
but in the counterclockwise direction (we completely neglect the
dynamical effects of the magnetic field).  In the next section, we
will consider the dominant classes of trajectory which contribute
to the current.  Before doing so, however, we pause to note that
in the limit $\nu \rightarrow 0, Z\rightarrow 1/2,$
Eq.~(\ref{eq:currentfinal}) reduces to the Ishii's
formula~\cite{REF:Ishii} for a wide clean SNS junction,

\begin{equation}
\label{eq:ishii} I_{\mathrm{tr.}}^{(0)}(\chi )=\sum_{n=1}^{\infty
}\frac{8v_{Fx}}{\beta v_{\mathrm{F}} \sinh \left( 2\pi l
n/\left( \beta v_{\mathrm{F}}\right) \right) } (-1)^{n+1}\sin
n\chi .
\end{equation}
Despite this similarity, our expression
Eq.~(\ref{eq:currentfinal}) generalizes Eq.~(\ref{eq:ishii}) by
incorporating normal reflection.  In particular, our expression is
not expected to yield a $2\Phi_0$-periodic pattern seen in the
theoretical results~\cite{REF:Barzykin},\cite{REF:Ledermann}
because it does not include, e.g., non-horizontal straight
trajectories, unless they are a part of a (closed) class A
trajectory.

\section{Summation over classes of trajectories}
\label{SEC:4}
In the present section we apply the results of the previous
section to calculate the contribution to the Josephson current due
to various types of quasiparticle trajectories.  We assume that
each such contribution has the form of the general expression
Eq.~(\ref{eq:currentfinal}) but with values of $\nu$, $l$ and
$\Omega$ taken from the geometrical properties of the associated
trajectory.

\begin{figure}[hbt]
\par
\epsfxsize=2.9in
\par
\par
\centerline{\epsfbox{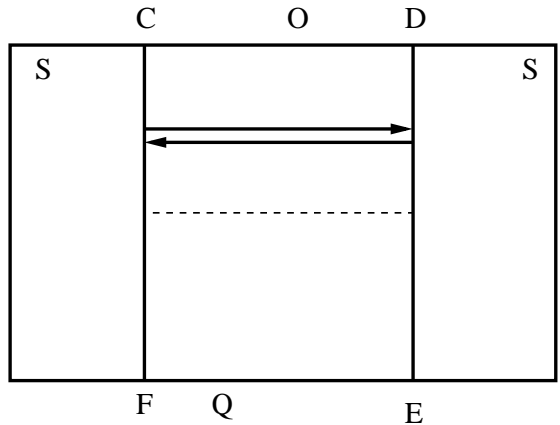}}
\par
\vskip0.50cm \caption{ Typical straight trajectory contributing to
the Josephson current. The intercept of the trajectory along the
segment $CF$  is the coordinate $y$ in Eq.~(\ref{eq:straight}). }
\label{fig:straight}
\end{figure}

  We begin by discussing the leading-order contribution
due to {\bf straight trajectories}, which undergo no reflections
from the sides of the device. Due to the fact that normal
reflection is a specular process, the trajectories must have
incoming angle $\theta = 0$, as depicted in
Fig.~\ref{fig:straight}.
It is easily seen that for them, the phase gains
[i.e.~Eqs.~(\ref{eq_nu},\ref{eq_Omega})] due to the magnetic field are $\nu =
0$ and $\Omega = -2y\phi/W$, where $\phi \equiv \Phi/\Phi_0$ with
$\Phi_0$ the flux quantum.  The length $l$ is given by $L$, and
$v_{Fx}= v_{\mathrm{F}}$.
Such trajectories are labeled by the coordinate $y$, given by
their intercept on the line segment $CF$ in
Fig.~\ref{fig:straight}.  To incorporate all such trajectories, we
integrate over $y$ from $-W/2$ to $W/2$, dividing by an overall
normalization factor $W$. Thus, we obtain for the current $I_{\rm
s}$ due to the straight trajectories
\begin{equation}
I_{\rm s} \approx \frac{4v_{\mathrm{F}}  x}{L \pi}
\int_{-\frac{1}{2}}^{\frac{1}{2}} dy \sum_{n=1}^{\infty} \frac{
(-1)^{n+1}}{\sinh nx} \sin n\Gamma(\chi -2y\phi),
\label{eq:straight}
\end{equation}
where for simplicity we have rescaled the dimensional coordinate
$y \rightarrow W y$ to make it dimensionless and have defined
$w\equiv W/L$ to be the dimensionless width of our sample. The
parameter $x\equiv 2\pi L/v_{\mathrm{F}} \beta$ is a dimensionless
measure of the temperature.

%


\begin{figure}[hbt]
\par
\epsfxsize=2.9in
\par
\par
\centerline{\epsfbox{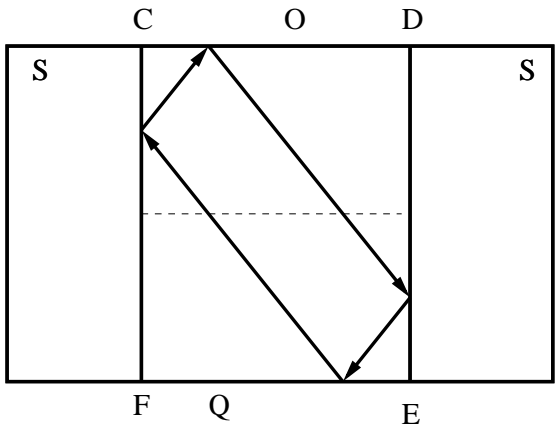}}
\par
\vskip0.50cm \caption{A typical bouncing trajectory.  The
intercept of the trajectory on the line segment $CF$ is the
coordinate $y$ in  Eq.~(\ref{eq:bounce}).
 }
\label{fig:two}
\end{figure}


Now consider the contribution due to class A (closed) {\bf
bouncing trajectories}, which bounce specularly {\em once} from
the {\em walls} of the device.
 The requirement
that the trajectory bounce {\it specularly} means, that there is
only one for each intercept $y$.   To compute this contribution,
we must calculate the values of $l$, $\nu$ and $\Omega$. These
are found geometrically to be
\begin{eqnarray}
\label{eq:nu}
l &=& L\sqrt{1+w^2} \\
\nu &=& 2\phi(y^2-\frac{1}{4})
\\
\Omega&=& 0,
\end{eqnarray}
where we have again rescaled $y\rightarrow yW$.
Summing over all such trajectories amounts to integrating over $y$
(and normalizing by an overall factor of $1/W$ as before) and
multiplying by an overall factor of $2$ to include
counterclockwise trajectories.  Thus, we find for the current
$I_{\rm b}$ due to the bouncing trajectories:
\begin{eqnarray}
&&I_{\rm b} \approx  \frac{8v_{\mathrm{F}} x}{\pi L\sqrt{1+w^2}}
 \int_{-\frac{1}{2}}^{\frac{1}{2}} dy
\sum_{n=1}^{\infty} \frac{  \cos(\pi \nu n)}{\sinh nx\sqrt{1+w^2}}
\nonumber \\
&&\qquad \qquad \times(-1)^{n+1}\sin n\Gamma(\chi).
\label{eq:bounce}
\end{eqnarray}
The generalization of Eq.~(\ref{eq:bounce}) to the case of
multiple reflections from the sides of the device is
straighforward.

\section{Results and discussion}

Having noted the various dominant contributions to the current, we
now turn to the calculation of the critical Josephson current
$\ic$ as a function of the applied flux.  This quantity is defined
in terms of the total Josephson current $I$ ($= I_{\rm s} + I_{\rm
b}$) as

\begin{equation}
\ic(\phi) \equiv {\rm max}\{ I(\chi,\phi)\}|_{0\leq \chi < 2\pi} .
\end{equation}

\begin{figure}[hbt]
\par
\epsfxsize=2.9in
\par
\par
\centerline{\epsfbox{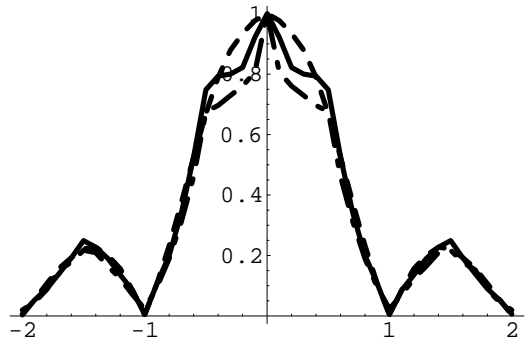}}
\par
\vskip0.50cm \caption{
 The evolution of the normalized critical current (computed numerically) as a function of
the external flux $\Phi/\Phi_0$ for the case of a wide junction.
The dot-dashed curve is  $Z=1$, the solid curve is $Z=2$ and the
dashed curve is $Z=10$. } \label{fig:wide}
\end{figure}

We begin by noting that  the contribution due to the bouncing
trajectories (i.e, Eq.~(\ref{eq:bounce})) is explicitly dependent
on the width-to-length ratio of the junction $w$ and indeed
vanishes for large $w$.  Thus, we can effectively isolate the
effect of the straight trajectories by examining a very wide
sample.  In Fig.~\ref{fig:wide}, we display numerical calculations
of $I_c(\phi)$
 for the case $w=50$, $Z =1,2,10$, and $x=0$ (i.e., $T=0$).
(All curves have been normalized to their values at $\phi = 0$.)


\begin{figure}[hbt]
\par
\epsfxsize=2.9in
\par
\par
\centerline{\epsfbox{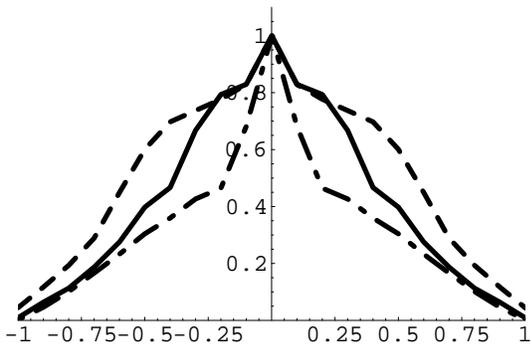}}
\par
\vskip0.50cm \caption{
 The evolution of the normalized critical current (computed numerically) as a function of
the external flux $\Phi/\Phi_0$ for the case of a narrow  junction
($w=1$), indicating a progressive sharpening of the Fraunhofer
diffraction pattern as $Z$ is increased. The dot-dashed curve is
$Z=10$, the solid curve is $Z=2$ and the dashed curve is $Z=1$. }
\label{fig:narrow}
\end{figure}

The pattern in Fig.~\ref{fig:wide} is $\Phi_0$-periodic (i.e.,
$\ic$ first vanishes at $\phi = \pm 1$), which is expected for a
wide junction~\cite{REF:Barzykin}. For a barrier height on the
order of the pair potential ($Z=1$, dot-dashed curve), there is
clearly a flattening of  $\ic$ below $\phi= \pm 1$ arising from
the energy-dependent amplitude for Andreev reflection. This
flattening (but not the sharp peak for $\phi \simeq 0$) is
reminiscent of the flattening effect seen in the 
JHAT experiment~\cite{REF:Takayanagi}.
This effect disappears for larger barrier heights, as exhibited by
the $Z=10$ curve (dashed line)
 in Fig.~\ref{fig:wide}. This shows approximate reverting to the
 Fraunhofer shape is to be expected, as the contribution
  of higher-order processes to the Josephson current
   is suppressed by increased normal reflection probability.
At smaller values of $w$, the increased importance of the bouncing
trajectories leads to somewhat different behavior of the
$\ic(\phi)$ curve as a function of $Z$. In Fig.~\ref{fig:narrow},
we display $\ic(\phi)$ for the case of a narrow junction (i.e.,
$w=1$) for the cases $Z = 1,2,10$ and $|\phi|<1$.  This curve is
also for the case of $T = x = 0$.
The $Z=1$ curve (dashed line) does exhibit flattening for small
values of $\phi$. In contrast to the wide junctions, increasing
the barrier height leads to a sharpening of the $\ic$ curve near
$\phi = 0$, as seen in the $Z=10$ curve (dash-dot line).


\section{Concluding remarks}

In this paper we have studied the effects of a finite tunneling
barriers and restricted geometries on the Josephson effect in a
clean junction. We derived
 closed formulas for the density of states and Josephson current in the system,
 which generalize the known results to the case of BTK normal scattering on
 NS interfaces, which allow an intuitive insight in the mechanisms of non-Fraunhofer
 diffraction patterns in mesoscopic SNS junctions and can be used in complex geometries.

 As an application of the technique, we have calculated the critical
 Josephson current $\ic$ in a rectangular SNS junction for various values
of the junction width and the tunneling barrier height $Z$.  As
expected, the deviation of $\ic(\phi)$ from the case of a clean
wide junction is most pronounced at small $\phi$, leading to a
flattening effect, reminiscent of the experimental results of
JHAT~\cite{REF:Takayanagi}. Nevertheless there remain open
questions.

First, our calculations do not provide a quantitative match. More
importantly, they do not reproduce the fine oscillations
superimposed on the flattened first peak, seen in the experiment of
Ref.~\onlinecite{REF:Takayanagi}. A possible source of this effect can
be due to the closed trajectories with multiple reflection points,
not taken into account here. The fine structure in the density of
states in SN billiard structures (i.e., so-called Andreev
billiards) arises from such trajectories~\cite{REF:Adagideli}, 
in accordance with the Gutzwiller
formula, which make this scenario at least plausible.
Finally, we neglected the elastic scattering inside the normal 
region, which was significant in Ref.~\onlinecite{REF:Takayanagi}.  
Generally, it would keep $l\neq\bar{l}$, so that that calculations up to
Eq.~(\ref{eq:matrix}) would hold, but the following formulas would be 
modified.


\noindent

\textsl{Acknowledgments\/}: We gratefully acknowledge stimulating
and enjoyable discussions with M.H.S.Amin, A.Golubov, Y.Harada,
A.Maassen van den Brink,  and H.Takayanagi. AZ thanks J.P. Hilton
for valuable comments on the manuscript. This work was supported
by NSERC.





\begin{thebibliography}{99}
\bibitem{note:scattering}
The second assumption is generally justified in S-2DEG-S 
structures, where the elastic scattering length $l_e$
can exceed 1 $\mu$m~\cite{REF:Heida}, but does not 
hold in the JHAT experiment~\cite{REF:Takayanagi},
where $l_e = .2\mu$m is on the order of the length of
the normal region ($0.22$ to $0.72 \mu$ m), but 
not its width ($4.2 \mu$m).
\end{thebibliography}
\end{document}